\documentclass{appolb}
\usepackage{graphicx}
\usepackage{lineno}
\begin{document}
\title{Search for the Chiral Effect using isobar collisions and BES-II data from STAR%
\thanks{Presented at Quark Matter 2022}%
}
\author{Yu Hu$^{1,2}$ (for the STAR Collaborations)
\address{1. Fudan University; 2. Brookhaven National Laboratory}
\\[3mm]
}
\maketitle
\begin{abstract}
In these proceedings we discuss the recent precision measurements of charge separation difference between Ru+Ru and Zr+Zr collisions at $\sqrt{s_{\rm NN}}=200$ GeV by STAR collaboration. The measurements indicate that the magnitude of the difference in the charge separation attributable to the magnetic fields between the two systems is smaller than previously expected. We also present charge separation measurements on the Chiral Magnetic Effect search from the RHIC BES-II experiment using the Event Plane Detectors (EPD) from Au+Au collisions at $\sqrt{s_{\rm NN}} =$ 27 GeV. 
\end{abstract}
  
\section{Introduction}
One of the major interests in the field of high energy physics is finding the experimental signatures of the local $CP$ violation in the strong interaction. It has been predicted that the changing of quark chirality is allowed in the QCD medium created in relativistic collisions through topological transitions. In the heavy-ion collision, with the strongest magnetic field  produced in nature, manifestations of such an effect are possible. An electric current is generated as a result of the imbalance of left-handed and right-handed quarks along with the magnetic field direction -- a phenomenon known as the Chiral Magnetic Effect (CME)~\cite{Kharzeev:1998kz}. 

There have been many experimental works to search for the  evidence of CME in the past two decades~\cite{Kharzeev:2015znc}. The CME sensitive $\gamma$ observable used for such searches is defined as~\cite{Voloshin:2004vk},
\begin{equation}
    \gamma^{\alpha,\beta}\equiv\langle \cos(\phi^{\alpha}+\phi^{\beta}-2\Psi_{2}) \rangle.
    \label{eq:dgamma}
\end{equation}
It is designed to measure the correlations of two charged particles, $\alpha$ and $\beta$, with respect to the event plane $\Psi_2$. As $\Psi_2$ is nearly perpendicular to the magnetic field, the $\gamma$ is expected to be sensitive to possible CME signal. 
To eliminate the charge-independent correlation background, driven by the global momentum conservation, the difference between opposite sign $\gamma$ ($\gamma_{OS}$) and the same sign $\gamma$ ($\gamma_{SS}$), $\Delta\gamma$, becomes the quantity of interest. Although a non-zero $\Delta\gamma$ has been observed at both RHIC and LHC energies, conclusive evidence of CME requires careful consideration of the charge-dependent backgrounds. Although many techniques have been developed over the years, disentangling the signal and background from the non-zero measurements is challenging. One expects the observed charge separation measured by $\Delta\gamma$ to have contributions as follows:  
\begin{equation}
    \Delta\gamma = \Delta\gamma^{CME} + k\frac{v_{2}}{N} + \Delta\gamma^{non-flow}.   
    \label{eq:dgamma-exp}
\end{equation}
Here $\Delta\gamma^{CME}$ is the signal term, $k\frac{v_{2}}{N}$ and $\Delta\gamma^{non-flow}$ are two background contributions from flow and non-flow, respectively. The magnetic field difference between the isobar collisions is expected to lead to a difference in the $\Delta\gamma^{CME}$ term. Thus, if two systems have similar flow-/non-flow-driven backgrounds, we  have the best chance to measure the signal difference. That is the basis of the isobar collisions designed and performed at RHIC~\cite{Voloshin:2010ut,STAR:2021mii,STAR:2019bjg}. 

\section{CME search in isobar collisions}
STAR used the isobar collision systems, Ruthenium+Ruthenium ($^{96}_{44}$Ru + $^{96}_{44}$Ru) and Zirconium+Zirconium ($^{96}_{40}$Zr + $^{96}_{40}$Zr). Among these isobar species, it has been argued that the magnetic field squared is $15\%$~\cite{Huang:2017azw} larger in Ru+Ru collisions, and the flow-driven background difference approximates $4\%$~\cite{Schenke:2019ruo}. We would expect to see $5\sigma$ difference when the background level in $\Delta\gamma$ is less than $80\%$ with 1.2 billion events in the two isobar systems. In the end, approximate 2 billion minimum-bias events for both species were collected by the STAR experiment~\cite{STAR:2021mii}.

\begin{figure}[htb]
    \centering
    \includegraphics[width=12.5cm]{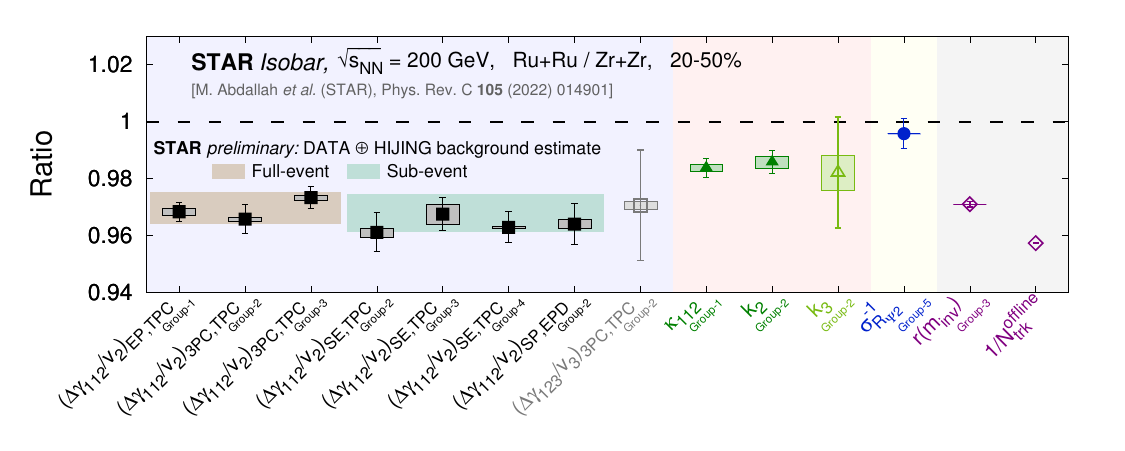}
    \caption{The compilation of post-blinding results and the current understanding of the non-flow contribution. The background estimation for full-event and sub-event methods are shown in brown band and green band, respectively.}
    \label{fig:isobar_post_bg}
\end{figure}

To minimize the unconscious biases, a blind analysis was proposed and applied~\cite{STAR:2019bjg}. In addition, five different teams inside the STAR collaboration participated in the blind analysis. A compilation of the results from the blind analysis is presented in Fig.~\ref{fig:isobar_post_bg}. 
The predefined CME criteria for the isobar blind analysis are:
\begin{equation}
    \frac{(\Delta\gamma_{112}/v_{2})^{Ru+Ru}}{(\Delta\gamma_{112}/v_{2})^{Zr+Zr}} > 1,
    \label{eq:Dg_basic}
\end{equation}
\begin{equation}
    \frac{(\Delta\gamma_{112}/v_{2})^{Ru+Ru}}{(\Delta\gamma_{112}/v_{2})^{Zr+Zr}} > \frac{(\Delta\gamma_{123}/v_{3})^{Ru+Ru}}{(\Delta\gamma_{123}/v_{3})^{Zr+Zr}},
    \label{eq:Dg_112_123}
\end{equation}
\begin{equation}
    \frac{(\Delta\gamma_{112}/v_{2})^{Ru+Ru}}{(\Delta\gamma_{112}/v_{2})^{Zr+Zr}} > \frac{(\Delta\delta)^{Ru+Ru}}{(\Delta\delta)^{Zr+Zr}}.
    \label{eq:Dg_Dd}
\end{equation}
Here the definitions for $\Delta\gamma_{112}/v_{2}$, $\Delta\gamma_{123}/v_{3}$, and $\Delta\delta$ can be found in Ref.~\cite{STAR:2021mii}.

In Fig~\ref{fig:isobar_post_bg}, the solid squares are the $\Delta\gamma_{112}/v_{2}$ ratio measurements between two systems using different methods. The open square point is the $\Delta\gamma_{123}/v_{3}$ ratio which can be regarded as a baseline for the $\Delta\gamma_{112}/v_{2}$ ratio as alluded in Eq.~\ref{eq:Dg_112_123}. We found that the isobar data are not compatible with the predefined CME criteria. The similar conclusion can also be made from the CME signal sensitive observables $\kappa$, $k_{n}$, and $\sigma^{-1}_{R_{\Psi_{2}}}$ measurements~\cite{STAR:2021mii}. The $\Delta\gamma/v_{2}$ ratios are below unity mainly driven by the multiplicity difference between the two isobars, although some deviations are observed between  $1/N^{offline}_{trk}$ and $\Delta\gamma/v_{2}$ ratios. The non-flow contribution needs to be considered to understand such a difference~\cite{yicheng:2022}. Therefore, in Fig~\ref{fig:isobar_post_bg} we present an estimate of the background after considering three additional sources of background on top of the naive inverse multiplicity ratio: the two-particle flowing cluster background, the two-particle non-flow, and the three-particle non-flow correlations. The first two terms are estimated using isobar data, the last term is estimated using the HIJING model~\cite{Feng:2021pgf,STAR:2021pwb}. The $\Delta\gamma/v_{2}$ measurements from the blind analysis of the isobar data are consistent with this new background estimate (shown by bands in Fig.~\ref{fig:isobar_post_bg}) within the the uncertainties.

\section{CME search at a lower energy with Au+Au collisions}
\begin{figure}
    \centering
    \includegraphics[width=10cm]{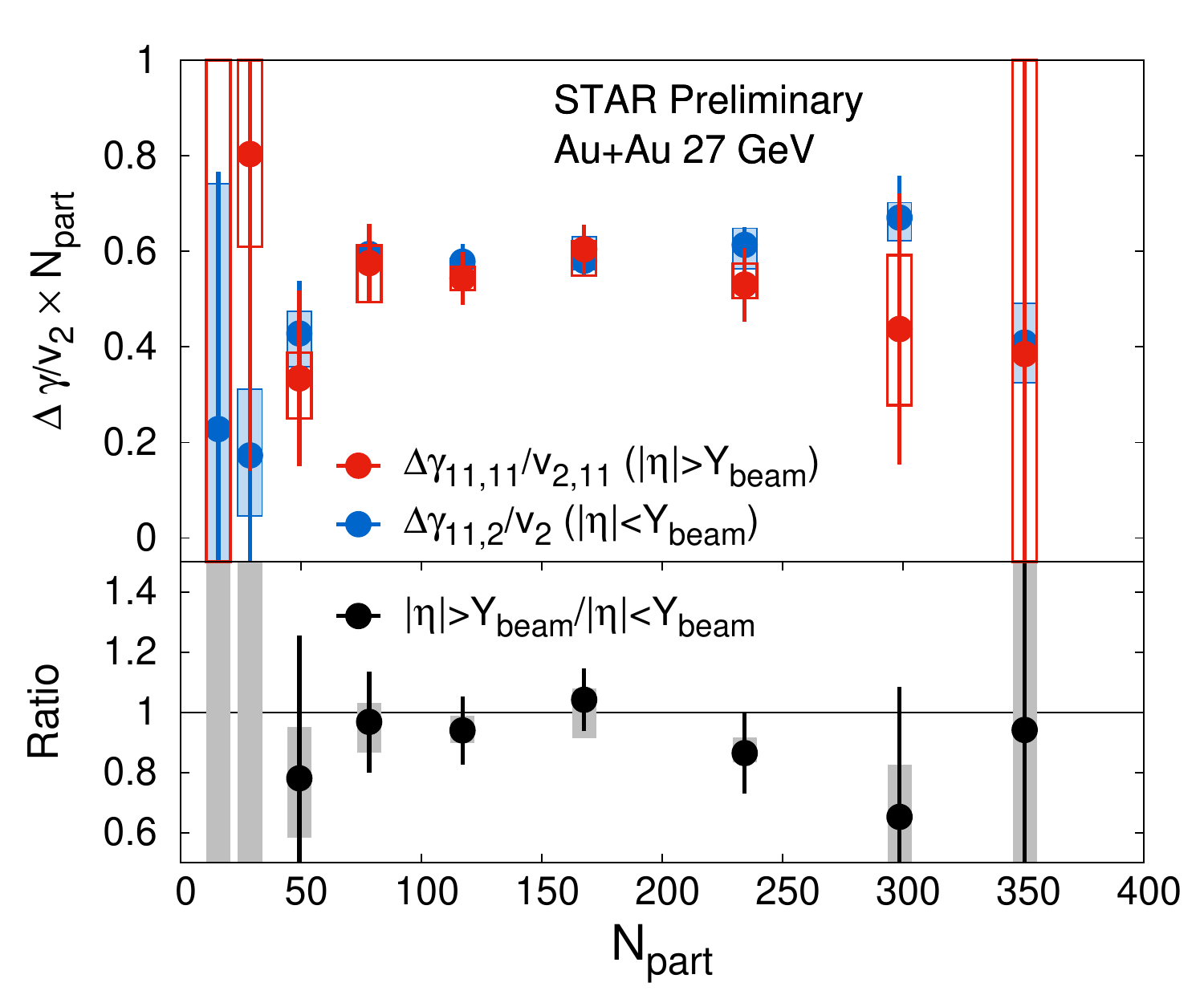}
    \caption{The charge separation measurements using the Event Plane Detectors at 27 GeV with Au+Au collisions. A CME driven correlation will drive the ratio in the lower panel larger than the unity~\cite{Hu:2021drc}.}
    \label{fig:27gev_epd}
\end{figure}
It is important to extend the CME search beyond the top RHIC energy because the prerequisites of the phenomenon have been argued to have a strong dependence on collision energy~\cite{Skokov:2009qp}. STAR has reported the charge separation measurements over a wide range of collision energies using the Beam Energy Scan I (BES-I) data in Au+Au collision~\cite{STAR:2014uiw}. Now with the newly installed Event Plane Detector (EPD) at STAR~\cite{Adams:2019fpo}, and ten times more statistics of the BES-II data give us a chance for higher precision measurements and an opportunity to better understand the signal and background in the CME measurements.

One unique promise of Au+Au $\sqrt{s_{NN}}=$ 27 GeV BES-II data collected by STAR experiment with the EPDs is as follows. The EPDs are located on both sides of Time Projection Chamber and cover the rapidity range of $2.1<|\eta|<5.1$. At $\sqrt{s_{NN}}=$ 27 GeV, the beam rapidity is $Y_{\rm beam}=$ 3.4 which falls in the middle of the EPD acceptance. The inner region of EPD detects spectator protons, whose directed flow signal has an opposite direction compared to the outer sectors that are dominated by hits from the participants or produced particles. Thus EPDs give us a new opportunity to measure the charge separation with respect to two different planes, the first order event plane ($\Psi_{1}$) enriched with spectator protons, and the second order event plane of the participants ($\Psi_{2}$).

Under the assumptions of purely flow-driven background scenario, the $\Delta\gamma/v_{2}$ measurements with respect to different planes should be similar. Any deviation of the double ratio of the quantity $\Delta\gamma/v_{2}(\Psi_1)/\Delta\gamma/v_{2}(\Psi_2)$ from unity would be interesting in the context of CME signal~\cite{Xu:2017qfs,Voloshin:2018qsm}. The preliminary results from STAR are shown in Fig.~\ref{fig:27gev_epd}. The double ratios measured with respect to $\Psi_1$ and $\Psi_2$ planes are consistent with each other -- indicating that the results are consistent with the scenario of background.

Besides the measurement using inner and outer EPD, a study using the event shape engineering technique~\cite{Milton:2021wku} has been reported to search for the CME signal at this conference. The background is significantly reduced with this approach. A quantitative investigation of the remaining background is needed for the measurement.

\section{Summary}
The blind analysis to search for CME signal using isobar data found that no predefined criteria are satisfied for the observation of CME.  The ongoing non-flow studies using the isobar data and HIJING model enable us to estimate an improved background baseline. The data are found to be consistent with such background estimate within the uncertainties. The high statistics BES-II data, the EPDs, and new techniques open new opportunities for the CME search at lower energies. 

\section{Acknowledgement}
Y. Hu is supported by the China Scholarship Council (CSC). This work was supported in part by the National Natural Science Foundation of China under contract No. 11835002.

\bibliographystyle{unsrt}
\bibliography{main}

\end{document}